\documentclass[a4paper,11pt]{article}
\usepackage{pos}

\newcommand{\be}{\begin{equation}}
\newcommand{\ee}{\end{equation}}
\newcommand{\ba}{\begin{eqnarray}}
\newcommand{\ea}{\end{eqnarray}}
\newcommand{\bi}{\begin{itemize}}
\newcommand{\ei}{\end{itemize}}

\newcommand{\<}{\langle} 
\renewcommand{\>}{\rangle}

\newcommand{\la}{\label}

\title{Deep inelastic scattering off quark-gluon plasma
  \\ and its photon emissivity}
 \ShortTitle{Deep inelastic scattering off quark-gluon plasma and its photon emissivity}

\author[a]{Marco C\`e}
\author[b]{Tim Harris}
\author*[c,d,e]{Harvey B.\ Meyer}
\author[c]{Arianna Toniato}
\author[c]{Csaba T\"or\"ok}

\affiliation[a]{Theoretical Physics Department, 
CERN, CH-1211 Geneva 23, Switzerland}

\affiliation[b]{School of Physics and Astronomy,
  University of Edinburgh, EH9 3JZ, UK}

\affiliation[c]{PRISMA$^+$ Cluster of Excellence \& Institut f\"ur Kernphysik, Johannes Gutenberg-Universit\"at Mainz,\\
Saarstr.\ 21, 55122 Mainz, Germany}

\affiliation[d]{Helmholtz Institut Mainz, Johannes Gutenberg-Universit\"at Mainz,
Saarstr.\ 21, 55122 Mainz, Germany}

\affiliation[e]{GSI Helmholtzzentrum f\"ur Schwerionenforschung, Planckstra\ss{}e 1, 64291, Darmstadt, Germany}

\emailAdd{meyerh@uni-mainz.de}
\abstract{
The photon emissivity of quark-gluon plasma probes the interactions in
the medium and differs qualitatively between a weakly coupled and a
strongly coupled plasma in the soft-photon regime.  The photon
emissivity is given by the product of kinematic factors and a spectral
function associated with the two-point correlator of the
electromagnetic current at lightlike kinematics.  A certain Euclidean
correlator at imaginary spatial momentum can be calculated in lattice
QCD and is given by an integral over the relevant spectral function at
lightlike kinematics.  I present a first exploratory lattice
calculation of this correlator.  Secondly, I show how Euclidean
correlators at imaginary spatial momenta can also be used to probe the
regime of deep inelastic scattering off quark-gluon plasma, which
reveals its parton distribution function.
\medskip

Preprint number: CERN-TH-2021-167
}

\FullConference{%
 The 38th International Symposium on Lattice Field Theory, LATTICE2021
  26th-30th July, 2021
  Zoom/Gather@Massachusetts Institute of Technology
}

\begin{document}
\maketitle

\section{Introduction}

In this presentation I will be concerned with the vector spectral functions of strong-interaction matter
at non-vanishing temperature, which can be defined as (in diag$(+,-,-,-)$ metric)
\be
\rho^{\mu\nu}(q) = \int d^4x\,e^{iq\cdot x} \, \frac{1}{Z}\sum_n e^{- E_n/T}
\<n| [j^\mu(x),j^\nu(0)]|n\>, \qquad Z = \sum_n e^{- E_n/T}.
\ee
Here $Z$ is the canonical partition function, and $j^\mu$ is a
conserved vector current, the case of the electromagnetic current
$j^\mu = \frac{2}{3}\bar u \gamma^\mu u - \frac{1}{3} \bar d
\gamma^\mu d - \frac{1}{3} \bar s \gamma^\mu s+ \dots$ being
particularly important for the phenomenology of heavy-ion collisions~\cite{Braun-Munzinger:2015hba,David:2019wpt}
and the physics of the early universe (see e.g.~\cite{Asaka:2006rw}), since it is via this current that the medium
interacts with photons. Therefore we will have this
case in mind in the following.

\begin{figure}
  \centerline{\includegraphics[width=0.6\textwidth]{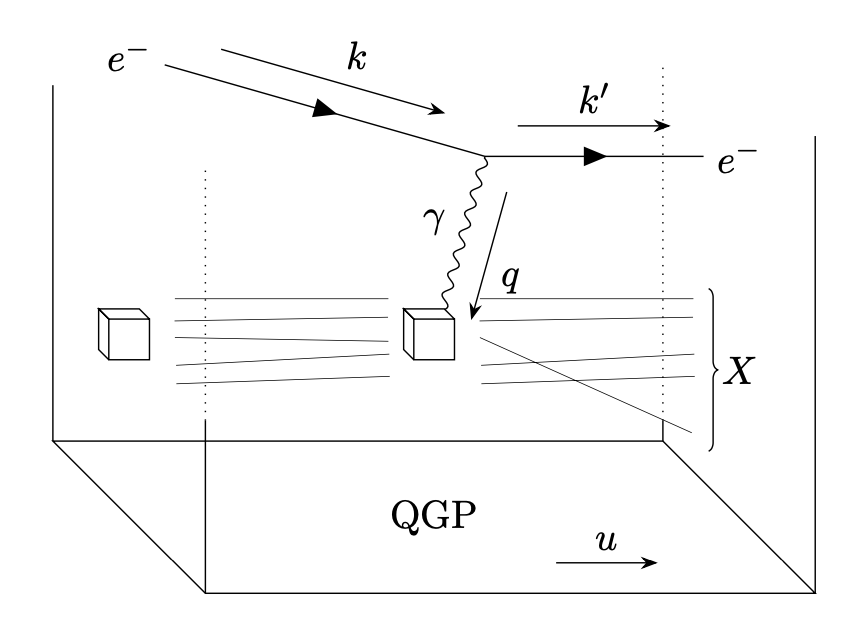}}
  \caption{\label{fig:illu}Scattering of a lepton on quark-gluon plasma at global thermal equilibrium moving with a four-velocity $u$ in the one-photon exchange approximation.
    The picture shows the interaction occurring with a cubic fluid cell, producing an unobserved QCD final-state $X$.}
\end{figure}

The components of $\rho^{\mu\nu}(q)$ can be parametrized by two
independent kinematic variables, which can be chosen to be the
virtuality $q^2$ and the photon energy $q^0$ in the rest frame of the
thermal medium.  At vanishing virtuality $q^2=0$, the spectral
functions describe the rate at which the medium emits photons (see
Eq.\ (\ref{eq:emissi}) below), and for positive virtuality, they
describe the production rate of lepton pairs via a timelike
photon~\cite{McLerran:1984ay}.  At spacelike virtualities, the vector
spectral functions measure the ability of the medium to convert the
energy stored in external electromagnetic fields into
heat~\cite{Ce:2020tmx}.  A conceivable way to create electromagnetic
fields with frequencies on the order of a GeV is to scatter a lepton
of energy $E$ on the medium. The process is illustrated in Fig.\ \ref{fig:illu}.
In the one-photon exchange approximation, the 
cross-section for scattering off the medium of volume $L^3$ reads
\be
\frac{d^2\sigma}{dE'd\Omega} = \frac{e^4\,L^3}{8\pi^2Q^4} \left(\frac{E'}{E}\right) \ell_{\mu\nu}W_>^{\mu\nu}(u,q),
\ee
in the rest frame of the medium, with $E'$ the final-state energy of the lepton and $\Omega$ the solid angle
of its outgoing momentum relative to the incident momentum~\cite{Harris:2020ijy}.
Here $\ell_{\mu\nu} $ is the (exactly known) leptonic tensor and  $W_>^{\mu\nu} $ the `hadronic' tensor,
names that we borrow from the literature on deep-inelastic scattering on the nucleon.
The hadronic tensor is defined as
\be
W_>^{\mu\nu}(u,q) = \frac{1}{4\pi\,Z} \sum_n e^{-\beta E_n} \int d^4x\; e^{iq\cdot x}\,\<n|j^\mu(x)\,j^\nu(0)|n\>,
\ee
with $u$ the four-velocity of the medium,
and is related by the Kubo--Martin-Schwinger relation to the spectral functions in the rest frame of the medium~\cite{Meyer:2011gj},
\be
W_>^{\mu\nu}(u,q) = \frac{1}{4\pi(1-e^{-\beta q^0})}\,\rho^{\mu\nu}(q^0,\vec q), \qquad  u=(1,\vec 0).
\ee
Particularly for studying the deep-inelastic scattering (DIS) limit, a suitable tensor decomposition is
\ba\la{eq:Wmunu_decomp}
W_>^{\mu\nu}(u,q) &=& F_1(u\cdot q,Q^2) \Big(-g^{\mu\nu} + \frac{q^\mu q^\nu}{q^2}\Big)
\nonumber\\ && + \frac{T}{u\cdot q} F_2(u\cdot q,Q^2)
\Big(u^\mu-(u\cdot q) \frac{q^\mu}{q^2}\Big) \Big(u^\nu-(u\cdot q) \frac{q^\nu}{q^2}\Big),
\ea
with $F_1$, $F_2$ being the structure functions of the medium.
We introduce the Bjorken variable
\be
 x = \frac{Q^2}{2T q^0},
\ee
in which the temperature $T$ represents the energy scale of the medium used here
to make $x$ dimensionless and $Q^2=-q^2$. Note that $x$ is not bounded from above by unity in the present context.

\section{ Probing the structure functions with Euclidean correlation functions}

\begin{figure}
  \centerline{\includegraphics[width=0.8\textwidth]{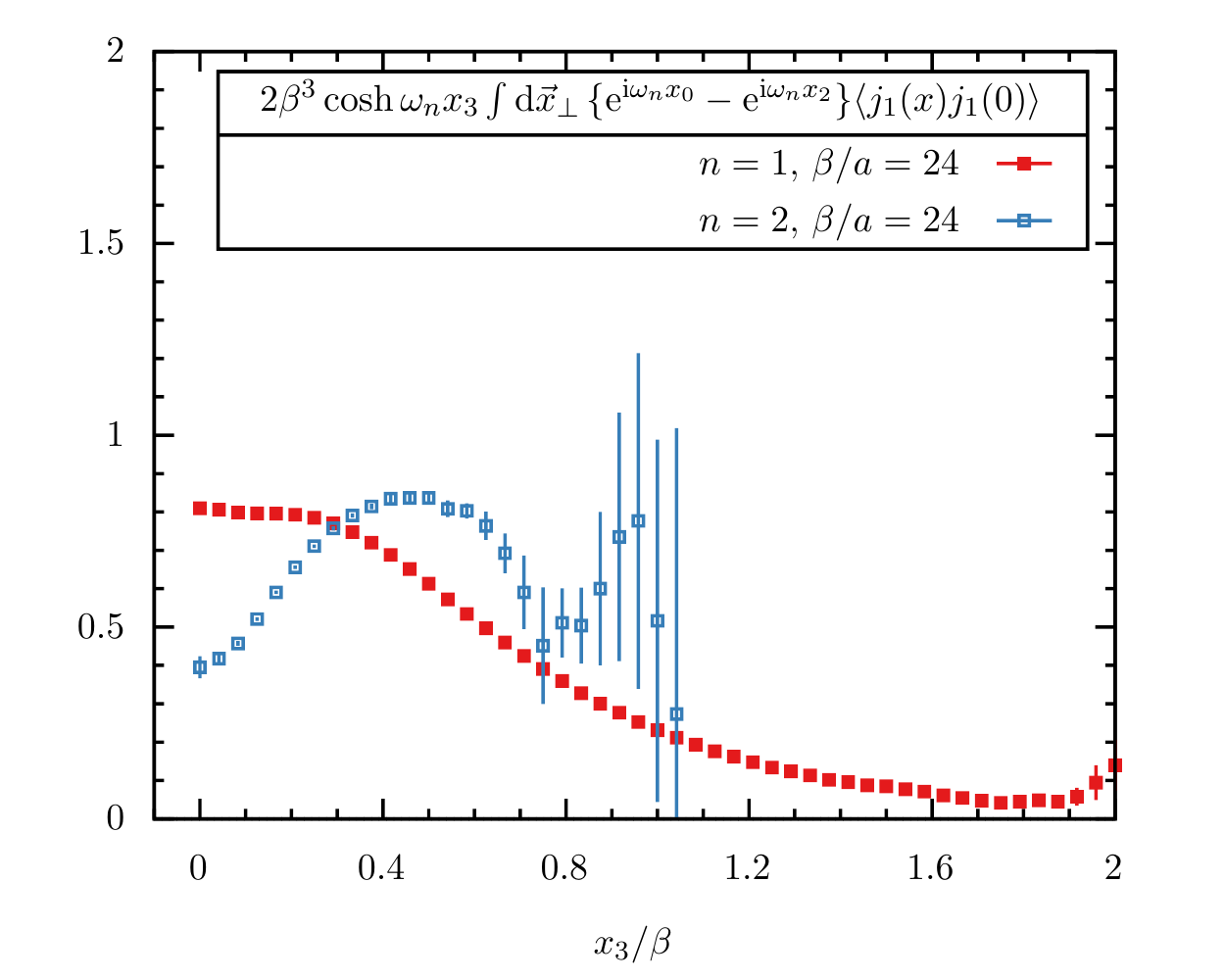}} % {fixv_sub_X7_ns64_integrand.pdf}}
  \caption{\label{fig:intgd} The $x_3$-integrand to obtain $H_E^T(\omega_n,0)$ based on Eq.\ (\ref{eq:H_Eesti}) for 
     the isovector current with quark charges ${\cal Q}_u=-{\cal Q}_d=1/\sqrt{2}$, after integrating over $d{\vec x}_\perp \equiv dx_0 dx_1 dx_2$.
     The area under both displayed correlators would be $T^2/2$ for free massless quarks~\cite{Meyer:2018xpt}.
    In general, the non-negativity of $\sigma^T$ implies that the area can be no smaller for the $n=2$ than for the $n=1$ correlator.
    The calculation is performed
    on a $24\times 96^3$ lattice QCD ensemble with two dynamical flavours of light quarks and $T\simeq 254\,$MeV, the X7 ensemble
    described in~\cite{Ce:2020tmx}. The displayed data is based on 1331 configurations, with 64 point sources used per configuration.
   At the largest $x_3$, the effect of the periodicity of this variable can be seen in the $n=1$ correlator.}
\end{figure}

We focus here on the spatially transverse channel represented by the structure function $F_1$ and 
define the corresponding spectral function
\be\la{eq:sigmaT} 
\sigma^{T} (q^0, Q^2) =  8\pi (1-e^{-\beta q^0}) F_1(q^0,Q^2),
\ee
viewed as a function of the photon energy and virtuality. Choose now a Matsubara frequency $\omega_n=2\pi n T$ and
a particular value of $Q^2\leq\omega_n^2$.
In~\cite{Harris:2020ijy}, it is shown that the $q^0$-dependence of this spectral function can be probed via the Euclidean correlator
\ba\la{eq:H_E_T}
H_E^{T} (\omega_n; Q^2) &=& -2\int_0^\beta dx_0\int d^3x\; e^{i\omega_n x_0 + x_3\sqrt{\omega_n^2-Q^2}}\; \big\< j_1(x)\,j_1(0)\big\>.
% H_E^{T} (\omega_n; Q^2) &=& 
% \int_0^\beta dx_0\int d^3x\; e^{\sqrt{\omega_n^2-Q^2}\hat{\vec q}\cdot \vec x + i\omega_n x_0}\;
% \Big({\txts\frac{1}{2}} \bigl(\delta_{ik} - \hat q_i \hat q_k \bigr)\; \< j_i(x)\,j_k(0)\>\Big).
%
%
% \nonumber\\
% H_E^{L} (\omega_n; Q^2) &=& \int d^4x\; e^{\sqrt{\omega_n^2-Q^2}\hat{\vec q}\cdot \vec x + i\omega_n x_0}\;
% \Big(\<j_0(x)\,j_0(0)\> + \hat q_i \hat q_k  \< j_i(x)\,j_k(0)\>\Big).
% \la{eq:H_E_L}
% \nonumber
\ea
This correlator contains a logarithmic ultraviolet divergence\footnote{An exception to this statement is the
case of $Q^2=0$, when the theory is regularized in a Lorentz-invariant way.}, which cancels in the difference
on the left-hand side of Eq.\ (\ref{eq:HEdisprel}). Indeed this difference would vanish in the vacuum, since
in the latter state $H_E^{T} $ only depends on $Q^2$.
In order to reach these conclusions, it is sufficient to Taylor-expand around $x=0$ the weight function
with which the position-space vector correlator is multiplied, taking into account that this correlator
is even under the coordinate transformation $x_0 \to -x_0$ as well as under $x_3\to -x_3$.
For $\omega_n^2\geq Q^2$, note the `imaginary spatial momentum' involved in computing the correlator $H_E^T$.
This correlator admits the fixed-virtuality dispersive representation
\be\la{eq:HEdisprel}
 H_E^{T} (\omega_n; Q^2) - H_E^{T} (\omega_r; Q^2) = 
\int_0^{\infty} \frac{ d q^0}{\pi} \: q^0 \: \sigma^{T} (q^0, Q^2) 
\Big[\frac{1}{(q^0)^2+\omega_n^2} -  \frac{1}{(q^0)^2+\omega_r^2} \Big]\: .
%
%  H_E^{T} (\omega_n; Q^2) - H_E^{T} (\omega_r; Q^2) = 
% \int_0^{\infty} \frac{ d \omega}{\pi} \: \omega \: \sigma^{T} (\omega, Q^2) 
% \Big[\frac{1}{\omega^2+\omega_n^2} -  \frac{1}{\omega^2+\omega_r^2} \Big]\: .
\ee
Since $\sigma^T(q^0,Q^2=0)\sim (q^0)^{1/2}$ at weak
coupling~\cite{CaronHuot:2006te}, one subtraction is expected to be
sufficient to guarantee convergence of the dispersion relation.  In
practice, one has the option to compute the left-hand side of
Eq.\ (\ref{eq:HEdisprel}) using at long distances $e^{x_3
  \sqrt{\hat\omega_n^2 - Q^2}}$ with $\hat\omega_n^2 = \omega_n^2 -
{\rm O}(a^2)$ slightly reduced in magnitude, in order to avoid
convergence issues in the infrared at non-zero lattice spacing.

% \bi
% \item Let $\sigma(\omega) \equiv \rho_T(\omega,|\vec q|=\omega)$ be the relevant spectral function proportional to the photon emission rate;
% \bigskip\item let $H_E(\omega_n)\equiv G_E(\omega_n,k=i\omega_n)$ the momentum-space Euclidean correlator with Matsubara frequency $\omega_n$
% and imaginary spatial momentum $k=i\omega_n$;
% \bigskip
% \item  once-subtracted dispersion relation:
% \ei
% \medskip

% \be
% H_E(\omega_n) - H_E(\omega_r) 
%  = \int_0^\infty \frac{d\omega}{\pi}\,{\omega}\,\sigma(\omega)
% \Big[ \frac{1}{\omega^2+\omega_n^2} -  \frac{1}{\omega^2+\omega_r^2}\Big], \quad n,r\neq 0.
% \ee
% \bigskip

Of special interest is the case of zero virtuality, since the differential photon emissivity of the medium is given by
\be\la{eq:emissi}
\frac{d\Gamma_\gamma}{dq^0} = \frac{\alpha}{\pi}\,\frac{q^0}{e^{\beta q^0}-1}\; \sigma^T(q^0,Q^2=0).
% \frac{d\Gamma_\gamma}{d\omega} = \frac{\alpha}{\pi}\,\frac{\omega}{e^{\beta\omega}-1}\; \sigma^T(\omega,Q^2=0).
\ee
At $Q^2=0$,  $H_E^{T}$ is ultraviolet-finite in a regularisation respecting Lorentz symmetry.
On the lattice however, it is necessary to perform a subtraction in order to obtain
the correct continuum limit for the correlator $H_E^T(\omega_n)$~\cite{Meyer:2018xpt}. For that purpose, we 
introduce here the estimator
\be\la{eq:H_Eesti} H_E^T(\omega_n,0)= -2 \int_0^\beta dx_0
\, \int d^3x\, (e^{i\omega_n x_0} - e^{i\omega_n x_2})\, e^{\omega_n  x_3}\; \big\<j_1(x) j_1(0)\big\>.
\ee
Note that the subtraction term
($e^{i\omega_n x_2}$) vanishes in the continuum thermal theory, but on the
lattice removes an ultraviolet divergence associated with the lack of
Lorentz symmetry at finite lattice spacing.
This is easiest to see from the fact that 
estimator (\ref{eq:H_Eesti}) of $H_E^T(\omega_n)$ automatically vanishes
on the $(T=0)$ infinite lattice.
With estimator (\ref{eq:H_Eesti}),
no cumbersome subtraction of vacuum correlators is needed, which would require costly simulations.
The integrand of Eq.\ (\ref{eq:H_Eesti})
in a lattice QCD calculation at $T\approx 250\,$MeV is illustrated in Fig.\ \ref{fig:intgd}.
Similar to the comment below Eq.~(\ref{eq:HEdisprel}), 
the exponential weight factor $e^{\omega_n  x_3}$ can be replaced by a factor with a slightly (O($a^2$)) reduced
exponent, for instance by substituting $\omega_n \to \frac{2}{a} \sin \frac{a\omega_n}{2} $.
Given that $H_E^T(0,0)=0$, we can write a simplied dispersion relation for the light-like case,
\be
H_E^{T} (\omega_n,0) = - \frac{\omega_n^2}{\pi} \int_0^\infty \frac{dq^0}{q^0}\; \frac{\sigma^{T} (q^0) }{(q^0)^2+\omega_n^2}\;.
\ee

\section{Thermal medium structure functions in the DIS regime}

\begin{figure}
  \centerline{\includegraphics[width=0.8\textwidth]{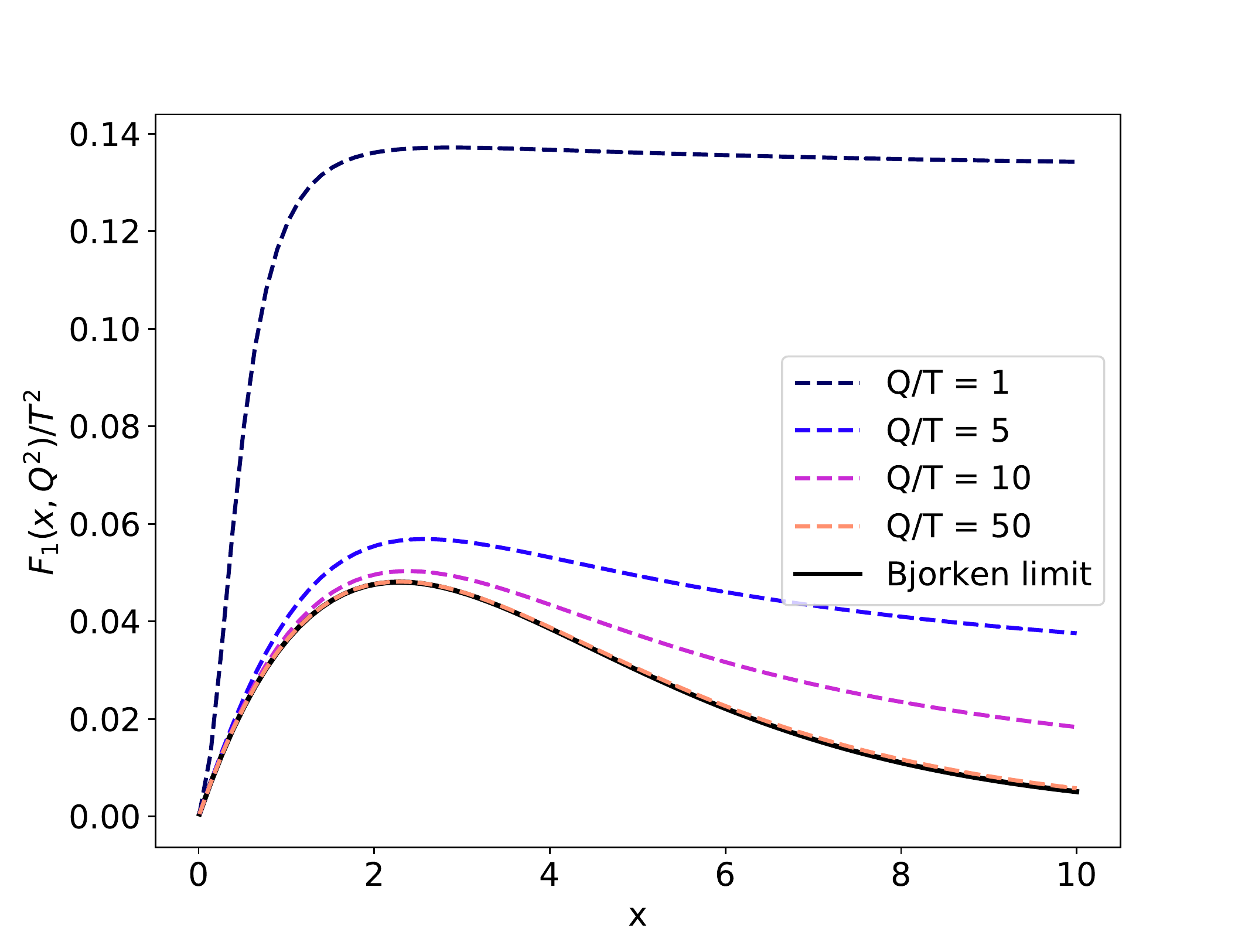}}
  \caption{\la{fig:freeF1} The structure function $F_1(x,Q^2)$ of the free plasma
    for different values of $Q^2$, together with the Bjorken limit Eq.~(\ref{eq:freeF1}).
    Here we have set $N_c=3$ and $\sum_f {\cal Q}_f^2=1$.
 }
\end{figure}

As in the case of the nucleon, one can show that the structure
functions admit a partonic interpretation in the DIS
limit~\cite{Harris:2020ijy}. With $u$ being the four-velocity of the
fluid, let $f_f(\xi)\, d\xi$ represent the number of partons of type
$f$ in the fluid cell that carry momentum $\xi \: T \: u$. This number
is of order the cell volume, $L^3$.
In the DIS limit, the structure function reads
\be\la{eq:F1f}
F_1(u\cdot q,Q^2) \stackrel{Q^2\to\infty}{\longrightarrow} \frac{1}{4L^3T} \sum_f {\cal Q}_f^2 \;f_f(x), 
\ee
with $x= Q^2/(2Tu\cdot q)$ kept fixed.
In words, $4\,F_1\cdot dx$ is the square-electric-charge weighted number of
partons carrying a momentum $xT$ times the fluid four-velocity $u^\mu$
per unit transverse area in a slab of fluid which in its rest frame
has thickness $1/T$ in the longitudinal direction.

In the theory of free quarks, in which the spectral functions can be computed analytically~\cite{Laine:2013vma},
one finds in the DIS limit
 \be
\lim_{Q^2\to\infty} F_1(x, Q^2) = 
\frac{ (\sum_f \mathcal Q_f^2) N_cT^2}{4\pi^2} x \log(1+e^{-x/2}),
\label{eq:freeF1}
\ee
thus showing that the parton distribution is proportional to $x \log(1+e^{-x/2})$,
and normalized such that the partons altogether
carry the entire momentum of the aforementioned fluid cell
as given by its rest-frame enthalpy,
% $(L^2/T)(e+p) u$, $(e+p)$ being the enthalpy densi
consistently with the expectation from ideal hydrodynamics~\cite{Harris:2020ijy}.
The approach to the DIS limit as a function of $x$ is illustrated in the free-quark theory
in figure~\ref{fig:freeF1}. One can see that it is only in the limit $Q^2\to\infty$ that the structure
 function acquires the  normalizable form of a probability distribution in the variable $x$.
This represents a qualitative difference with respect to ordinary DIS on the nucleon.

How can the DIS regime of the structure functions be addressed using lattice QCD?
Using the twist expansion of the product of two vector currents, one finds the moment sum rules
obeyed by the structure functions~\cite{Harris:2020ijy}, 
\be
\int_0^{\infty} d x \: x^{n-1} \: [F_1 (x, Q^2)]_{\textrm{leading-twist}} =  
\frac{1}{2}\sum_{f,j} {\mathcal Q_f^2} M_{fj} (Q, \tilde \mu) \langle O_{nj} \rangle \: , \quad n = 2,4,\dots \: ,
\label{eq:F1_sum_rule}
\ee
where the $M_{fj} $ are the coefficients parametrizing the operator mixing and
the $O_{nj}^{\mu_1 \dots \mu_n}$ are the usual twist-two operators of mass dimension $(n+2)$,
whose thermal expectation values are parametrized as follows,
\be
\langle O_{nj}^{\mu_1 \dots \mu_n} \rangle = T^n [u^{\mu_1} \dots u^{\mu_n}  - \mathrm{traces}]\langle O_{nj} \rangle \: .
\ee
The index $j$ runs over the quark flavours and the gluon operator.
The right-hand side of the moment sum rule is computable on the lattice for the lowest few values of $n$.
A difference with ordinary DIS is however that the leading-twist component of $F_1$ must be taken before
calculating the $x^{n-1}$ moment in Eq.\ (\ref{eq:F1_sum_rule}), as the integral otherwise does not converge,
as can be seen from Fig.\ \ref{fig:freeF1}.
Alternatively to computing the lowest $x$-moments of the structure functions,
the approach to the DIS limit can be studied by computing the correlators $H_E^{T} (\omega_n; Q^2)$
for a sequence of Matsubara frequencies $\omega_n$. Performing the change of variables
$x= Q^2/(2T q^0)$ in Eq.\ (\ref{eq:HEdisprel}), one obtains
\be
H_E^{T} (\omega_n; Q^2) - H_E^{T} (\omega_r; Q^2) = 
\int_0^{\infty} \frac{ d x}{\pi} \: x \: \hat\sigma^{T} (x, Q^2) 
\frac{a_r^2 - a_n^2}{(1 + a_n^2 x^2)(1 + a_r^2 x^2)}
\: ,
\ee
with $a_n = 2T\omega_n/Q^2$. Thus if $a_n$ and $a_r$ are kept fixed as $Q^2$ is varied,
the only $Q^2$ dependence on the left-hand side comes from the
spectral function $\hat\sigma^{T} (x, Q^2)\equiv \sigma^T(q^0,Q^2)$. The extent to which
it becomes independent of $Q^2$ in the Bjorken limit can thus be probed in this way.

\section{Conclusion}

Dispersion relations at fixed spacelike virtuality, rather than at
fixed spatial momentum, open up new perspectives on the thermal
spectral functions. Here we have shown that 
certain moments of the spectrum of emitted photons can be computed in lattice QCD
  without solving an inverse problem.
The general lesson is that when an analytic
continuation in one variable of a Euclidean correlation function
depending on several kinematic variables is performed, which
numerically amounts to solving an inverse problem, it plays an
important role which of these kinematic variables are kept fixed.

We have also seen that moments of the in-medium structure functions
can be computed in lattice QCD at fixed spacelike virtuality.  In this
kinematic regime, the structure functions describe the scattering of a
lepton on the medium, and, at sufficiently high virtuality in the
Bjorken regime, should reveal its quark-gluon constituents.  Such a
regime is however difficult to reach with the current simulation
techniques, due to the real exponential $e^{x_3\sqrt{\omega_n^2-Q^2}}$
emphasizing the large distances in the spatial directions.  One must
also remember that even in the theory of free quarks, the Bjorken
scaling is only reached after `higher-twist' (O($1/Q^2$)) contributions
are sufficiently suppressed.

\acknowledgments{This work was supported by the
European Research Council (ERC) under the European Union’s Horizon
2020 research and innovation program through Grant Agreement
No.\ 771971-SIMDAMA, as well as by the Deutsche
Forschungsgemeinschaft (DFG, German Research Foundation) 
through the Cluster of Excellence “Precision Physics, Fundamental
Interactions and Structure of Matter” (PRISMA+ EXC 2118/1) funded by
the DFG within the German Excellence strategy (Project ID 39083149).
The work of M.C.\ is supported by the European Union's Horizon
2020 research and innovation program
under the Marie Sk\l{}odowska-Curie Grant Agreement No.\ 843134-multiQCD.
T.H. is supported by UK STFC CG ST/P000630/1.
The generation of gauge configurations as well as
the computation of correlators was performed on the Clover and Himster2 platforms
at Helmholtz-Institut Mainz and on Mogon II at Johannes Gutenberg University Mainz.
The authors gratefully acknowledge the Gauss Centre for Supercomputing
e.V. (www.gauss-centre.eu) for funding project IMAMOM by providing
computing time through the John von Neumann Institute for Computing
(NIC) on the GCS Supercomputer JUWELS~\cite{JUWELS} at J\"ulich Supercomputing
Centre (JSC).}

\bibliographystyle{JHEP.bst}
\bibliography{/Users/harvey/BIBLIO/viscobib.bib}

% \begin{thebibliography}{99}
% \bibitem{...}
% ....

% \end{thebibliography}

\end{document}